\documentclass{easychair}
\usepackage{amssymb,amsmath} 
\usepackage{mathptmx} 

\usepackage{doc}
\usepackage{makeidx}

\usepackage{breakurl}             % Not needed if you use pdflatex only.
\usepackage{fancyhdr}
\usepackage{epsfig}
\usepackage{latexsym}
\usepackage{microtype}
\usepackage{hyperref}

\usepackage{times}
\usepackage{amssymb}
\usepackage{amsmath}
\usepackage{latexsym}
\usepackage{graphicx}

\usepackage[strings]{underscore} % For the doi problem in bibtex
\usepackage{mdwlist}
\usepackage{subfigure}
\usepackage{wasysym}
\usepackage{amsthm } %We are missing example and definition
\title{NTCCRT: A concurrent constraint framework for real-time interaction (extended version)}

\author{
Mauricio Toro\inst{1}
\and
Camilo Rueda\inst{2}
\and
Carlos Ag\'on\inst{3}
\and \\
G\'erard Assayag\inst{3}
}

\institute{
  Universidad Eafit,
  Medellin, Colombia\\
\and
   Pontificia Universidad Javeriana Cali,
   Cali, Colombia\\
\and
   IRCAM,
   Paris, France\\
 }

\title{\papertitle}

\begin{document}
%\DeclareGraphicsExtensions{.png,.jpg,.pdf} % used graphic file format for pdflatex
    %  \hyphenation{whe-re non-de-ter-mi-nis-ti-ca-lly Rue-da's }
\maketitle
% first paragraph in each section should not be in- 
% dented, but all other paragraphs should be.
%\begin{center}\textbf{{\Large Abstract}}\\\end{center} % NEW !!
\begin{abstract} %  150 words.
% We propose a new way to synchronize concurrent processes in signal processing
% languages such as Pure Data and Max. Pure Data and Max does not 
% provide primitives for synchronization. Although concurrent processes can
% be programmed extending these languages with C++, we argue that this approach is counter-intuitive for
% non-computer scientists and it is very difficult and time-
% demanding. We propose using the \texttt{ntcc} formalism to specify the
% synchronization %for these programming languages 
% and execute those
% specifications in real-time with our tool, \textit{Ntccrt}.
% % We provide some examples where this 
% %approach is very useful in music composition and signal processing.

%Motivation\\
\noindent
Writing multimedia interaction systems is not easy. Their concurrent processes
usually access shared resources in a non-deterministic order, often
leading to unpredictable behavior.
Using Pure Data (Pd) and Max/MSP is possible to program concurrency, however, it is difficult to synchronize processes based on multiple criteria.
Process calculi such as the Non-deterministic Timed Concurrent Constraint (ntcc) calculus, overcome that problem by representing multiple criteria as constraints.
%Most programming languages
%such as C++, do not provide a formal semantic to reason
%about the behavior of concurrent programs.
% In addition,
% concurrent visual programming, usually based on process calculi,
% makes the power of concurrency available for a wider range of people.
% Writing programs
%that way lead often to bugs, only detected in execution time.
% why user level?????? motivate
%Problem\\
% In fact, it is possible to write 
% concurrent programs in them using a threading API and 
% writing an external (usually in C++).
%Approach\\
We propose using our framework Ntccrt to manage concurrency in Pd and Max. Ntccrt is a real-time capable interpreter for ntcc.
% Ntcc can trigger actions based on
% multiple criteria, easily represented by entailment of constraints.
%Ntcc is a formalism for concurrency providing a formal semantic and
%verification procedures. 
% In Ntccrt, ntcc models are compiled as 
% an external (binary plugin) for Pd or Max.
%Results\\ %Conclusions\\%
%\textit{Ntccrt} allows the development multimedia interaction systems 
%graphically and execute them in Pd and Max.
Using Ntccrt externals (binary plugins) in Pd we ran models
for machine improvisation and signal processing.
%, taking advantage
%of audio and MIDI processing tools provided by Pd.
%In the future, we want a graphical interface to represent ntcc processes directly on Pd.

\end{abstract}
%\\
%\\
%\noindent
%\textbf{Keywords: Max/MSP, Pure Data, NTCC, constraints, GECODE, concurrency} 
%\keyword{Keywords: Max/MSP, Pure Data, NTCC, constraints, GECODE, concurrency}
\section{Introduction} 
\label{sec:introduction}
Multimedia interaction systems --inherently concurrent-- can be modeled using concurrent process calculi. Process calculi are useful 
to describe formally the behavior of concurrent systems, and to prove properties about the systems.
Process calculi has been applied to the modeling of 
 spatially-explicit ecological systems \cite{PT13, TPSK14, PTA13, mean-field-techreport} and interactive multimedia systems
 \cite{is-chapter,tdcr14,ntccrt,cc-chapter,torophd,torobsc,Toro-Bermudez10,Toro15,ArandaAOPRTV09,tdcc12,toro-report09,tdc10,tdcb10,tororeport} .

% Writing programs
% Writing correct concurrent systems,
% using textual programming languages to program concurrent systems is 
% intended to be only for specialized users and programmers.
% Concurrent visual programming, usually based on concurrency formalisms \cite{cordial},
% makes the power of concurrency available for a wider range of people. 

%\subsubsection{Ntcc, a process calculus}
For instance, using \texttt{ntcc} \cite{ntcc}, we can model reactive systems with synchronous, asynchronous and/or non-deterministic behavior.
\texttt{Ntcc} and extensions have been used 
to model interactive systems such as: an audio processing framework
\cite{audiontcc},  machine improvisation \cite{Rueda06a},
\cite{pntcc}, \cite{rtcc}, and interactive scores \cite{AADR06},
\cite{rtcc}.
% The novelty of this approach is specifying the synchronization in declarative way. % Explain, what is declarative?
%\subsubsection{Not suitable for real-time, the problem}

Although there are three interpreters for \texttt{ntcc}, they 
are not suitable for real-time (RT) interaction.
It means that they are not able to interact with the user without letting him experience noticeable delays in the interaction.
%\subsubsection{What is real-time?}
% RT interaction means that the user does not experience noticeable delays in the 
% interaction.
%\subsubsection{Real-time systems written in c++}
%\subsubsection{Pd and Max, the solution}

On the other hand, we can program RT systems for multimedia interaction and signal processing using C++. Unfortunately, using C++ requires long development time. To overcome that problem, programming languages such as Pu-re Data (Pd) \cite{pd} and Max/MSP \cite{max}, provide a graphical interface to program RT systems and several APIs for concurrent programming. 
%In addition, Pd and Max are among the most used software in the computer music community.

\subsection{The problem}
%\subsubsection{Threading apis and message passing}
%\subsubsection{difficult to write complex concurrent programs, the problem}
Although
Pd and Max support concurrency, it is a hard
task to trigger or halt the execution of a process
based on multiple criteria.   
%\subsubsection{what are complex concurrent programs?}
% It is difficult because synchronization in Pd and Max
% is usually programmed by sending and receiving control messages
% among \textit{graphical objects}. 

Using Pd or Max, it is hard to express: ``process $A$ is going to do an action $B$ until a condition $C$ is satisfied'', when condition $C$ is a complex condition resulting from many other processes' actions. Such condition would be hard to express (and even harder to modify afterwards) using the graphical patch paradigm.
%  To achieve such behavior, we need to send messages to the process $A$, telling it to execute an action $B$. When the condition $C$ is satisfied, we stop
% sending messages to $A$.
For instance, condition $C$ can be a conjunction of these criteria: 
(1) The user has played on a certain tonality, (2) has played the chord G7, and (3) played the note F\# among the
last four.

% \textbf{Paragraph 2: What is the specific problem considered in this paper? This paragraph narrows down the topic area of the paper. In the first paragraph you have established general context and importance. Here you establish specific context and background.}

% The idea is not creating a new
% programming language based on $Ntcc$. We are interested in creating 
% interpreters for $Ntcc$ specifications, taking advantage of
% all the libraries (e.g. signal processing, GUI, midi controls)
% provided by different programming languages.

% \textbf{Paragraph 3: "In this paper, we show that ...". This is the key paragraph in the intro - you summarize, in one paragraph, what are the main contributions of your paper given the context you have established in paragraphs 1 and 2. What is the general approach taken? Why are the specific results significant? This paragraph must be really really good. If you can't "sell" your work at a high level in a paragraph in the intro, then you are in trouble. As a reader or reviewer, this is the paragraph that I always look for, and read very carefully.

% You should think about how to structure this one or two paragraph summary of what your paper is all about. If there are two or three main results, then you might consider itemizing them with bullets or in test (e.g., "First, ..."). If the results fall broadly into two categories, you can bring out that distinction here. For example, "Our results are both theoretical and applied in nature. (two sentences follow, one each on theory and application)"}

\subsection{Our solution}
%\subsubsection{Ntccrt for Max and Pd, the solution}
% In this work, we propose the  calculus to manage concurrency
% in data-flow programs. $Ntcc$ is a formalism s. Additionally, it provides multiple
% agents who can reason about partial information represented by constraints. 

Using \texttt{ntcc}, we can represent the complex condition $C$ presented above
as the conjunction of constraints ($c_1 \wedge c_2 \wedge c_3$).
Each constraint (i.e., mathematical condition) represents a criterion. In addition, each criterion
can be represented declaratively. For instance, the criterion (2) can
be represented by the constraint ``G7 is on the set of played chords'' ($G_7 \in PlayedChords$).

For that reason, we propose using \texttt{ntcc} to manage concurrency in Pd and Max,
executing \texttt{ntcc} models on \\\textit{Ntccrt}\footnote{This research was partially founded by the REACT project, sponsored by Colciencias.}. %\textit{Ntccrt} is our RT capable interpreter for \texttt{ntcc}.
On \textit{Ntccrt}, \texttt{ntcc} models can be automatically compiled as 
an \textit{external} (i.e., a binary plugin) for Pd or Max. 

Additionally, the externals can be specified textually using Common Lisp or graphically using OpenMusic \cite{Bresson05b}.
We argue that concurrent visual programming, usually based on process calculi (such as Cordial \cite{cordial}),
makes the power of concurrency available for a wider range of users.
% $Ntccrt$ uses the Generic Constraints Development Environment  ($Gecode$ \cite{fastprop}) to manage constraints and concurrency.
%  For instance, we provide an interface for OpenMusic \cite{Bresson05b}, allowing the user to write, graphically, \texttt{ntcc} specifications and translating them to either: stand-alone programs interacting with the real-time library Midishare \cite{midishare}, or externals for Pd or Max.

% In order to fix that inconvenient, we built an interpreter in the C++ programming language, capable of real-time ($Ntccrt$). 

% \textbf{Paragraph 5: "The remainder of this paper is structured as follows..." Give the reader a roadmap for the rest of the paper. Avoid redundant phrasing, "In Section 2, In section 3, ... In Section 4, ... " etc.}
\subsection{Contributions}
Our framework \textit{Ntccrt}  (http://ntccrt.sourceforge.net) is composed by the following components. The \texttt{ntcc} interpreter written in C++ and interfaces for both Common Lisp and OpenMusic. In addition, we provide the implementation of two real-life applications.
%  In addition, in this paper, we explain the design and implementation of the interpreter, its two interfaces and two real-life applications.
% % \begin{itemize}
% \item 
% \item 
% \item 
% \item 
% \end{itemize}

\subsection{Structure of the paper}

The remainder of this paper is structured as follows. Section 2 intuitively explains the semantic of \texttt{ntcc} agents and gives some examples of simple \texttt{ntcc} processes modeling multimedia interaction. Section 3 explains related work on \texttt{ntcc} interpreters and threading APIs available for Pd and Max.  Section 4 discusses two applications of \textit{Ntccrt} to model a multimedia interaction and a signal processing system. Section 5 explains our results. Finally, section 6 gives concluding remarks and proposes future works.

\section{The ntcc calculus}
A family of process calculi is Concurrent Constraint Programming (CCP) \cite{ccp}, where a system is modeled in terms of variables and constraints over some variables. Furthermore, there are agents reasoning about partial information (by the means of constraints) about the system variables contained on a common \textit{store}. 

CCP is based on the idea of a \textit{constraint system}. A constraint system includes a set of (basic) constraints and a relation (i.e., entailment relation $\models$) to deduce a constraint based on the information supplied by other constraints.
A CCP system usually includes several constraint systems for different variable types. 
 There are constraint systems for different variable types such as sets, trees, graphs and natural numbers. A constraint system providing arithmetic relations over natural numbers is known as Finite Domain (FD). For instance, using a FD constraint system we can deduce the constraint $pitch \neq 60$ from the constraints $pitch > 40$ and $pitch < 59$.

Although we can choose an appropriate constraint system to model any problem, 
in CCP it is not possible to delete nor change information accumulated in the store. For that reason, it is difficult to perceive a notion of discrete time, useful to model reactive systems (e.g., machine improvisation) communicating with an environment.

% A CCP calculus used to model multimedia interaction systems is the \textit{Non-deterministic Timed Concurrent Constraint (\texttt{ntcc} \cite{ntcc})}\cite{ntcc} calculus, where we can model reactive systems with synchronous, asynchronous and/or non- deterministic processes.

\texttt{Ntcc} introduces to CCP the notion of discrete time as a sequence of \textit{time-units}. Each time-unit starts with a \textit{store}  (possibly empty) supplied by the environment, then \texttt{ntcc} executes all processes scheduled for that time-unit.
In contrast to CCP, in \texttt{ntcc}, variables changing values along time can be modeled explicitly.
In \texttt{ntcc}, we can have a variable $x$ taking different values on each time-unit. To model that
in CCP, we would have to create a new variable $x_i$ each time we change the value of $x$.

For instance, a system that plays sequentially the notes of the C major chord can be modeled in \texttt{ntcc} as ``in the first time-unit, let $pitch=C$; in the second time-unit, let $pitch=E$; and in the third time-unit, let $pitch=G$''. Using CCP, we would represent it as ``let $pitch_1=C$, let $pitch_2=E$, and let $pitch_3=G$''.

% Opposed to the CCP model, in $Ntcc$ we
%can model variables changing through time, because they can change values from a time-unit to another.
%This semantic is really useful to reason about multimedia interactive systems according to Rueda and Valencia
%\cite{rueda-formalizing}.

% NEW OJO OJO
 %The computational agents of NTCC are summarized in table \ref{tab:ntccagents}. 
Following, we give some examples of how the computational agents of \texttt{ntcc} can be used with a FD constraint system.
%  The operational semantic of all \texttt{ntcc} agents can be found in Appendix \ref{operational} and 
A summary can be found in table \ref{tab:ntccagents}.

Using the
``tell'', it is possible to add constraints such as $\textbf{tell} (pitch_1 = 60)$, meaning that $pitch_1$ 
must be equal to 60 or $\textbf{tell} (60 < pitch_2 < 100)$, meaning that $pitch_2$  is an integer between 60 and 100. 

The ``when'' can be used to describe how the system reacts to different events. For instance, 
$\textbf{when}$ $pitch_1 = C \wedge pitch_2 = E \wedge pitch_3 = G$ $\textbf{do}$   $\textbf{tell} (CMayor = true)$ is 
a process reacting as soon as the pitch sequence C, E, G  has been played, adding the constraint 
$CMayor = true$ to the store in the current time-unit.
% (represented as 48, 52, 55 in Musical Instrument Digital Interface (MIDI) notation). 

\begin{table}
  \begin{center}   
     \begin{tabular}{|l l|}
\hline
Agent & Meaning\\
\hline
\textbf{tell} $(c)$ & Adds $c$ to the current store\\
\textbf{when} $(c)$ \textbf{do} $A$  & If $c$ holds now run $A$\\
\textbf{local} $(x)$ \textbf{in} $P$  & Runs $P$ with local variable $x$\\
$A$ $\|$ $B$ & Parallel composition \\
\textbf{next} $A$ & Runs  $A$ at the next time-unit \\
\textbf{unless} $(c)$ \textbf{next} $A$  & Unless $c$ can be inferred now, run $A$ \\
$\sum \limits_{i \in I}$ \textbf{when} $(c_{i})$ \textbf{do} $P_{i}$  & Chooses  $P_{i}$ s.t. $(c_{i})$ holds \\
*$P$ & Delays P indefinitely (not forever)\\
!$P$ & Executes P each time-unit\\
\hline
\end{tabular}    
    \caption{\texttt{Ntcc} agents}
    \label{tab:ntccagents}
  \end{center}
\end{table}

Parallel composition allows us to represent concurrent processes. For instance,  
$\textbf{tell}$ $(pitch_1 = 62)$ $\|$ $\textbf{when}$ $60 \leq pitch_1 < 72$ $\textbf{do}$ $\textbf{tell}$ $(Instrument = 1)$ 
is a process telling the store that $pitch_1$ is 62 and concurrently assigning the instrument to one, since $pitch_1$ is in first octave.
%The number one represents the acoustic piano in Musical Instrument Digital Interface (MIDI) notation.% (fig. \ref{fig:tellwhenparallel}).% (see in figure \ref{fig:tellwhenparallel}).

The ``next''  is useful
when we want to model variables changing through time. For instance, $ \textbf{when}$ $(pitch_1 = 60)$  $\textbf{do}$ $\textbf{next}$  $\textbf{tell}$ $(pitch_1 <> 60)$, means that if $pitch_1$ is equal to 60 in the current time-unit,  it will be different from 60 in the next time-unit.

% \begin{figure}[!h]
%   \begin{center}
%     \mbox{
%        \scalebox{0.6}{\includegraphics[width=\columnwidth]{images/tellwhenpar.pdf}}  
 
%       }
%     \caption{\textbf{Tell}, \textbf{when}, and \textbf{parallel} agents in \textit{Ntcc}}
%     \label{fig:tellwhenparallel}
%   \end{center}
% \end{figure}

% I.E = es decir , e.g. por ejemplo
The ``unless'' is useful to model systems reacting when a condition is not satisfied or it cannot be deduced from
the store. For instance, $\textbf{unless}$ $(pitch_1 = 60)$  $\textbf{next}$ $\textbf{tell}$ $(lastpitch \\<> 60)$ reacts when $pitch_1 = 60$ is false or when $pitch_1 = 60$ cannot be deduced from the store (e.g., $pitch_1$ was not played in the current time-unit), telling the store in the next time-unit that $lastpitch$ is not 60. % (see figure \ref{fig:unless}).

% \begin{figure}[!h]
%   \begin{center}
%     \mbox{
%        \scalebox{0.6}{\includegraphics[width=\columnwidth]{images/unless.pdf}}  
 
%       }
%     \caption{Unless agent in \texttt{ntcc}}
%     \label{fig:unless}
%   \end{center}
% \end{figure}

The ``star'' (\textbf{*}) may be used to delay the end of a process indefinitely, but not forever. For instance, $* \textbf{tell}$ $(End = true)$. 

The ``bang'' ($!$)  executes a certain process in every $time$-$unit$ after its execution. For instance, $!$$\textbf{tell}$ $(C_4 = 60)$. 

The \textbf{$\sum$} is used to model non-deterministic choices.
For instance, $!$ $\sum_{i \in \{48,52,55\}}$ \textbf{when} $i \in PlayedPitches$
\textbf{do} $\textbf{tell}$ \\$(pitch = i)$ models a system where each
time-unit, it chooses a note among the notes played previously that belongs to the C major chord.

Finally, a basic recursion can be defined in \texttt{ntcc} with the form $q(x) \overset{def}{=} P_q$, where $q$ is the process name and
$P_q$ is restricted to call $q$ at most once and such call must be within the scope of a ``next''.
The reason of using ``next'' is that \texttt{ntcc} does not allow 
recursion within a time-unit. Recursion is used to model
iteration and recursive definitions. For instance, using this basic
recursion, it is possible to write a function to compute the factorial function.

\section{Related work}
% \textbf{Paragraph 4: At a high level what are the differences in what you are doing, and what others have done? Keep this at a high level, you can refer to a future section where specific details and differences will be given. But it is important for the reader to know at a high level, what is new about this work compared to other work in the area.}
In this section, we present related work about concurrency support for Pd and Max,  and available \texttt{ntcc} interpreters.

\subsection{Writing concurrent programs on Pd and Max}
% The graphic environment facilitates the programming of
% interactive multimedia applications for %ARGUE WHY
% !!!!!!!!!!!!!!!!!!!!!!!!!!!!!!!!!!!!
% Human computer interaction !!! EXPLAIN
% ____________
%--------------------
% non -computer scientists according to the principles of Human-computer interaction. Examples of these programming languages are
% \textit{Pure Data (Pd \cite{pd})} and Max \cite{max} developed by Miller
% Puckette. Puckette's languages have been used for real-time synthesis and hardware
% control in several compositions and concerts.
%(http://www.harvestworks.org/maxreel/).

%\subsection{The problem}
% A computation in a data flow program starts by receiving an input. After that, the input goes
% through multiple \textit{graphical objects}. Each of them transforms the input into new messages, audio, or video signals. These \textit{graphical objects} can receive an input and transform it at any time. For that reason, data flow languages are inherently concurrent.

%\subsection{Overview of other solutions}
% Neither Max nor Pd provide \textit{graphical objects} to synchronize the chopsticks and the philosophers.
% In fact, creating a \textit{graphical object} for that purpose posses many difficulties to make 
% it work for an arbitrary number of philosophers.
% THis is not true. Explain further !!!!!!!
% further further
% further further
To program concurrent
applications on Max and Pd, we can use their message passing APIs. We can also create externals in C++.
In fact, we can use any existing threading API for C++ to write externals for both, Pd and Max. There is also a native API
for Max 5 SDK.
%  An external to synchronize the philosophers and the chopsticks can be made using a threading or co-routines API provided by C++.
Another way to write an external is using the Flext library. Flext provides a unique interface to write, in the C++ language,  externals dealing with both, Pd and Max.
% In \textit{Ntccrt}, we use Flext to generate portable \textit{Ntccrt} externals.
%CONNECTION ! \\

%CONNECTION ! \\
% There is another way, besides message passing and writing externals, to program concurrent applications on Max and Pd. This approach is the one used by the improvisation software Omax
% \cite{omax}. This software is composed by two modules. One module is
% in charge of signal processing (written in Max) and the other one is
% in charge of concurrency control and style learning (written in
% OpenMusic). The concurrency control is made using 
% Lisp processes (medium-weight threads for Common Lisp).

\subsection{Ntcc interpreters}
There are three interpreters available for \texttt{ntcc}: Lman \cite{lman} used as a framework to program Lego$^{TM}$ robots, NtccSim \cite{ntccsim} used to model and verify properties of biological systems,  
 and 
Rueda's interpreter \cite{Rueda06a} for multimedia interaction. 

% Since the development of $Lman$ back in 2003, multiple attempts 
% to run \texttt{ntcc} models of multimedia interaction systems have
% been made. The first attempt was made by Mu$\tilde{n}$oz and Hurtado. 
The first attempt to execute a multimedia interaction \\\texttt{ntcc} model was made by the authors of Lman in 2003.
They ran a \texttt{ntcc} model to play a sequence of
pitches with fixed durations in Lman. Recently, in
2006, Rueda et al. ran ``A Concurrent Constraint Factor Oracle Model for Music Improvisation'' ($Ccfomi$) on Rueda's
interpreter \cite{Rueda06a}. 

Both, $Lman$ and \textit{Rueda's interpreter} ran
the model giving the expected output. However, they were not
capable of executing multimedia interaction systems in real-time.

\section{Our framework: Ntccrt}
\textit{Ntccrt} is our framework to specify and execute \texttt{ntcc} models.% for Pd or Max.
% The first interpreter built was $Lman$  \cite{lman}, composed by an abstract machine written in the C language, a compiler and a visual
% language. It only supports $Fd$ constraint system and it was used to control Lego RCX robots.
% After $Lman$, $NtccSim$ came up . It was developed in Mozart-OZ \cite{MOZ2004}. It supports
% $Fd$ and real intervals \footnote{Real intervals (XRI) are constraint
%   systems used to represent real numbers.} as constraint systems and it was used to simulate
% biological systems \cite{biontcc}. Recently, Avispa and Ircam wrote \textit{Rueda's interpreter} \cite{Rueda06a} in Common Lisp. It supports
% $Fd$, finite sets ($Fs$), and rational trees \cite{rational} constraint systems and it
% was used to run a \textit{Concurrent Constraints Factor Oracle Model for
% Music Improvisation (CCFOMI)}. Unfortunately,
% none of them is able to achieve real-time in multimedia interaction.
% %real-time? on-line? off-line !!!!!!!!!!!!
% % what are you talking about?
% % explain
%What is \textit{Ntccrt}? \\

\subsection{Design of Ntccrt}
Our first version of \textit{Ntccrt} allowed us to specify \texttt{ntcc} models in C++ and execute them as stand-alone programs.
 Current version offers the possibility to specify a \texttt{ntcc}
model on either Lisp, Openmusic or C++. It is also possible to
execute \texttt{ntcc} models as a stand-alone program or as an
external object for Pd or Max.

In addition to its portability, $Ntccrt$ was carefully designed to
support Finite Domain, Finite Sets and Rational Trees constraint systems. Those constraint
systems can be used to represents complex data structures 
(e.g., automata and graphs) commonly used in computer music. 

\textit{Ntccrt} works on two modes, one for writing the models and another one for executing those models.
\subsubsection{Developing mode}
\begin{figure}[!h]
  \begin{center}
    \mbox{
      % \scalebox{0.6}
{\includegraphics[width=\columnwidth]{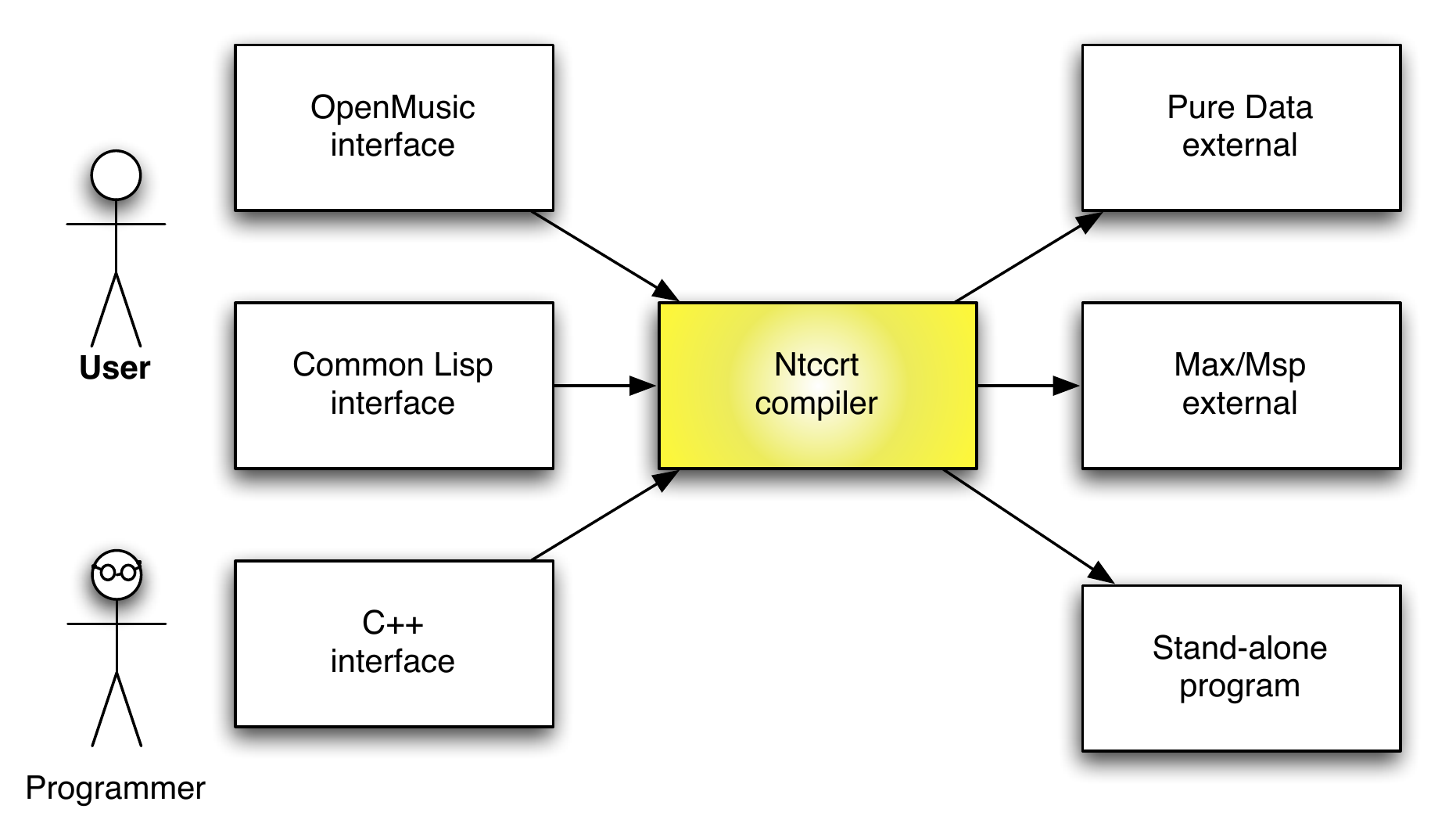}}  
 
      }
    \caption{\textit{Ntccrt}: Developing mode}
    \label{fig:developingmode}
  \end{center}
\end{figure}

In order to write a \texttt{ntcc} model in \textit{Ntccrt}, the users may
write them directly in C++, using a parser that takes Common Lisp macros
 or writing a graphical ``patch'' in OpenMusic. Using either of these representations, it is possible to generate a stand-alone program or
an external (fig \ref{fig:developingmode}).

%\subsection{Parsing}
%To make an interface for OpenMusic, first, we developed a Lisp parser using
% The Lisp interface is based on Common Lisp macros used to
%  write a \texttt{ntcc} model in Lisp syntax and translate
% it to C++ code. Lisp macros extend Lisp 
% syntax to give special meaning to characters reserved for users for this purpose. Those macros also automatically compiled a \texttt{ntcc} program.

% On the other hand, the OpenMusic interface uses OpenMusic methods 
% to represent \texttt{ntcc} processes. Openmusic methods are a graphical representation using the Common Lisp Object System (CLOS). 
% Those methods can be placed in a graphical ``patch'' (provided by OpenMusic) and the evaluation of the patch leads to the compilation of a \textit{Ntccrt} program.

%\section{Generating binary plugins for Pd and Max/Msp}

% %\subsection{Pd MAx Common Lisp and OpenMusic}
% Although we developed a portable, generic, and real-time capable
% interpreter for $Ntcc$; we still had a problem. In order to write a \texttt{ntcc}
% specification, it was necessary to write code in C++ and then
% compiling it. This was clearly counter-intuitive for
% non-computer scientists. For that reason, we developed a parser on top
% of $OpenMusic$, where both computer scientists and musicians, can
% write \texttt{ntcc} specifications in a graphical way. Every specification is automatically
%  compiled as an stand-alone application using \textit{Midishare} or as an external for $Pd$ or $Max$.

\subsubsection{Execution mode}
To execute a \textit{Ntccrt} program, we can proceed in two different ways. We can create a stand-alone program or we can create an external for either Pd or Max. An advantage of using the externals lies on using control signals and the message passing API provided by Pd and Max to synchronize any \textit{graphical object} with the \textit{Ntccrt} external.

% \begin{figure}[!h]
%   \begin{center}
%     \mbox{
%      %  \scalebox{0.6}
% {\includegraphics[width=\columnwidth]{images/executemode.pdf}}  
 
%       }
%     \caption{\textit{Ntccrt}: Interaction mode}
%     \label{fig:executemode}
%   \end{center}
% \end{figure}

%\subsection{Input/output of MIDI streams}
To handle Musical Instrument Digital Interface (MIDI) streams we
use the predefined functions in Pd or Max to process MIDI.
Then, we connect the output of those functions to the
\textit{Ntccrt} external. We also provide an interface for Midishare \cite{midishare}, useful when
running stand-alone programs.% (fig. \ref{fig:executemode}). % CITE midishare in the intro

% Another consideration, when designing $Ntccrt$, was to make
% it portable to different operative systems (those supported by
% $Gecode$) and to different programming languages (C++, Lisp,
% OpenMusic, Pd, Max). Furthermore, it was required that $Ntccrt$ was capable of interacting in real-time.

% In order to achieve those requirements, we developed $Ntccrt$ in the C++ language, for portability with different operative systems and programming languages (since most of the languages are extensible using C++ code). Furthermore we use the Generic Constraints Development Environment ($Gecode$ \cite{gecode}) library to represent all those constraint systems and to control the concurrency. 
\subsection{Implementation of Ntccrt}
\textit{Ntccrt} is written in C++ and it uses Flext to generate the externals for either Max or Pd, and Gecode \cite{fastprop} for constraint solving and concurrency control.
Gecode is an efficient constraint solving library, providing efficient propagators (narrowing operators reducing the set of possible values for some variables). 
The basic principle of \textit{Ntccrt} is encoding the ``when'', $\sum$ and ``tell'' 
processes as Gecode propagators.
The other processes are simulated by storing them into queues for each time-unit.

Although Gecode was designed to solve combinatorial problems, Toro found out in  \cite{tororeport} that writing the ``when'' and the $\sum$ processes as propagators, Gecode can manage all the
concurrency needed to represent \texttt{ntcc}.
Following, we explain the encoding of the ``tell'' and the ``when''.

To represent the ``tell'', we define a super class $Tell$. For \textit{Ntccrt}, we provide three subclasses to represent these processes: \textbf{tell} ($a=b$), \textbf{tell} ($a \in B$), and \textbf{tell} ($a > b$).
 %(see figure \ref{fig:tells}).
 Other kind of ``tells'' can be easily defined by inheriting
from the $Tell$ superclass and declaring an $Execute$ method.

\begin{figure}[!h]
  \begin{center}
    %\mbox{
       %\scalebox{0.7}
{\includegraphics[width=0.48\columnwidth]{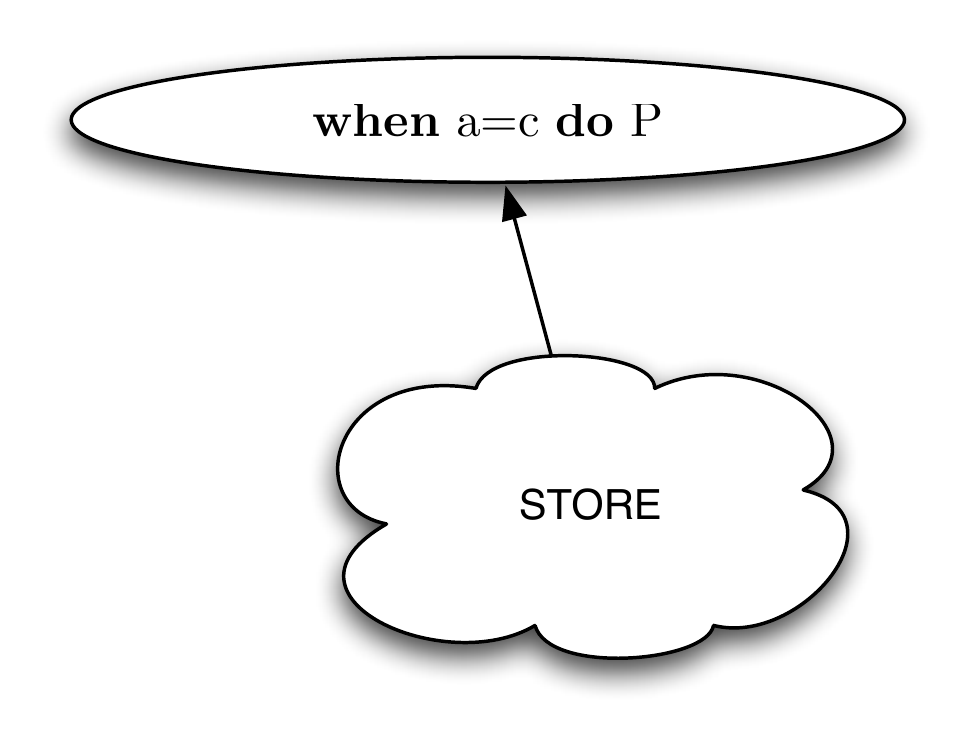}} \ \ %\quad
      % \scalebox{0.7}
{\includegraphics[width=0.48\columnwidth]{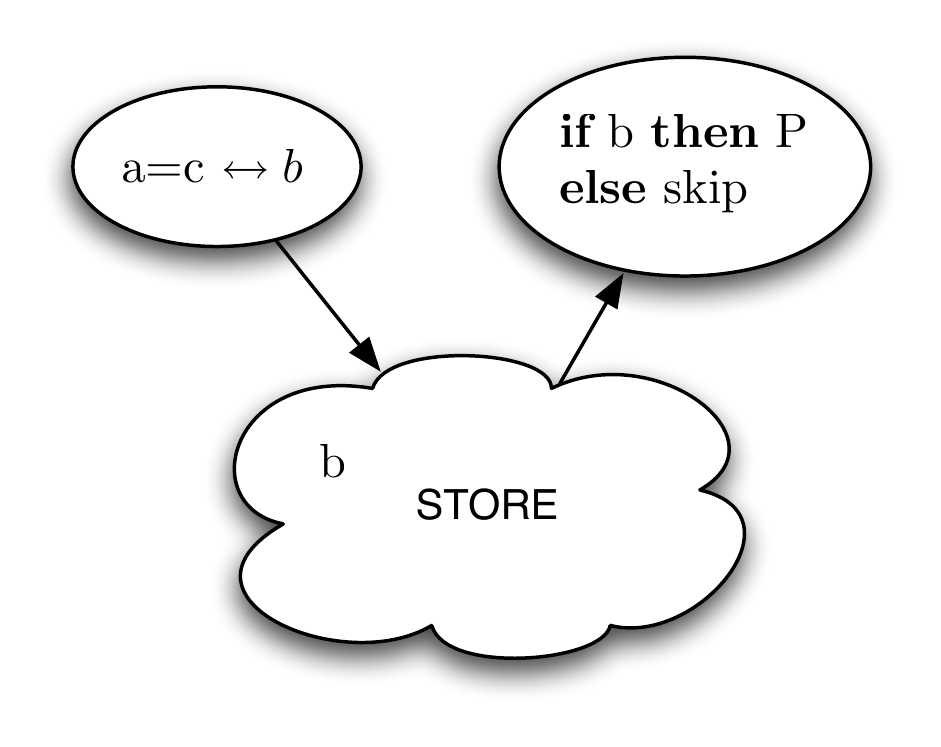}} 
     % }
    \caption{Example of the \textit{When propagator}}
    \label{fig:whenprop}
  \end{center}
\end{figure}

We have a \textit{When propagator} for the ``when'' and a \textit{When} class for calling the propagator.
A process
\textbf{when} $C$ \textbf{do} $P$ is represented by two propagators:
$C \leftrightarrow b$ (a reified propagator for the constraint $C$) and \textbf{if} $b$ \textbf{then} $P$ \textbf{else} $skip$ (the \textit{When  propagator}).
The \textit{When propagator} checks the value of $b$. If the value of $b$ is true, it calls the \textit{Execute} method
of $P$. Otherwise, it does not take any action. Figure \ref{fig:whenprop} shows how to encode the process \textbf{when} $a=c$ \textbf{do} $P$ using our \textit{When propagator}.

\section{Applications}

We selected two real-life applications to show the relevance of using
\textit{Ntccrt} externals in Pd. 
% They dining philosophers problem,
%  gives an intuition of how to model a simple synchronization problem
% in \texttt{ntcc}. 
 \textit{Ccfomi} shows us how we can use \textit{Ntccrt} to interact
in real-time with a human player. Finally, a signal processing application
 shows us how a \textit{Ntccrt external} can send control signals
to trigger signal processing filters.

\subsection{ Machine Improvisation}

% Musical improvisation is the spontaneous creative process of making music while it is being performed. To use a linguistic analogy, improvisation is like speaking or having a conversation as opposed to reciting a written text. 

Machine improvisation usually
consider building representations of 
music, either by explicit coding of rules or applying machine learning
methods.% For machine improvisation it is necessary to 
An interactive machine improvisation system capable of real-time perform two
activities concurrently: Stylistic learning and Stylistic simulation.

Rueda et al. define in \cite{Rueda06a}, Stylistic learning as the process of applying machine learning methods to musical sequences in order to capture salient musical features and organize these features into a model. On the other hand, Stylistic simulation as the process of producing musical sequences stylistically consistent with the learned material.

A machine improvisation system using \texttt{ntcc} is \textit{Ccfomi}. \textit{Ccfomi} executes both phases concurrently, uses \texttt{ntcc} to synchronize both phases
of the improvisation, and uses the \textit{Factor Oracle} ($FO$) to store the information of the
learned sequences. 

$FO$ is a finite state
automaton constructed in linear time and space. It has two kind of
transitions (\textit{links}). \textit{Factor links} are going forward and following them is
possible to recognize at least all the factors from a
sequence. \textit{Suffix links} are going backwards and they connect
repeated patterns of the sequence. Further formal definitions about $FO$ can be found in
\cite{Allauzen99factor}. 
% $FO$ has been used in the last decade in multiple models for music improvisation \cite{Assayag07a},\cite{cont08a},\cite{Rueda06a} because of its low complexity and because there is an incremental algorithm to build it.

Following, we give a brief description of \textit{Ccfomi} taken from \cite{Rueda06a}. \textit{Ccfomi} is divided in three subsystems: learning (ADD), improvisation (IMPROV) and playing 
(PLAYER) running concurrently. In addition, there is a synchronization
process (SYNC) in charge of synchronization.

\textit{Ccfomi} has three kind of variables to represent the partially built $FO$
 automaton: Variables $from_{k}$ are the set of labels of all currently existing
 \textit{factor links} going forward from $k$. Variables $S_{i}$ are \textit{suffix links} 
 from each state $i$ and variable  $\delta_{k,\sigma_{i}}$ give the state reached from $k$
 by following a \textit{factor link} labeled $\sigma_{i}$.

% The variables $from_k$ and  $\delta_{k,\sigma_{i}}$ are modelled as
% rational trees, allowing us to add elements to them each time unit.
% For example, with the constraints $cons(A,B)$, $cons(B,C)$, and $cons(C,D)$
% we can have a list of three elements [$A,B,C$,\_] and then we can add more
% elements, adding constraints to the variable $D$. 
% % CITE the rational trees !! Ya estaba

In our implementation of \textit{Ccfomi}, the variables $from_k$ and  $\delta_{k,\sigma_{i}}$ are modeled as
infinite rational trees \cite{rational} with unary branching.
That way, we can add new elements to $from$ and $\delta$ dynamically.
%, allowing us to add elements to them, each time-unit. 
%Infinite rational trees have infinite size. However, they only contain a 
%finite number of distinct sub-trees. 

Rational trees have been subject of multiple researches
to construct a constraint system based on them.
Using this constraint system is possible to post the constraints \\$cons(c,nil,B)$, $cons(b,B,C)$, $cons(a,C,D)$ to 
model a list of three elements [$a,b,c$].

Following, we explain some \textit{Ccfomi} processes.
The $ADD$ process (specified in \cite{Rueda06a}) is in charge of building the $FO$  by creating the \textit{factor links} and the \textit{suffix links}. This
process models the learning phase.

The learning and the simulation phase must
work concurrently. In order to achieve that, it is required that the simulation phase only takes place
once the subgraph is completely built. The $SYNC$ process is in charge of doing the synchronization
between the simulation and the learning \\phase to preserve that
property. 

Synchronizing both phases is greatly simplified by the use of constraints. When a variable has no value, the ``when'' processes depending on it are blocked. Therefore, the $SYNC$ process is ``waiting'' until $go$ is greater or equal than one. It means that the $PLAYER$ process has played the note $i$ and the $ADD$ process can add a new symbol to the $FO$. The condition $S_{i-1} \geq 0 $ is because the first \textit{suffix link} of the FO is equal to -1 and it cannot be followed in the simulation phase. \\

$SYNC_{i}$ $\overset{def}{=}$  \\
\hspace*{35pt}\textbf{when} $S_{i-1} \geq -1 \wedge go \geq i$ \textbf{do}\\
\hspace*{55pt} ($ADD_{i}$ $\|$ \textbf{next} $SYNC_{i+1}$) \\
\hspace*{30pt}$\|$ \textbf{unless} $S_{i-1} \geq -1 \wedge go \geq i$ \textbf{next} $SYNC_{i})$ \\

The $PLAYER$ (specified in \cite{Rueda06a}) process simulates a human
player. It decides, non-deterministically, each time-unit between
playing a note or not. When running this model in Pd, we replace this
process by receiving an input (e.g., a MIDI input) from the environment.

The improvisation process $IMPROV$  starts from state $k$ and probabilistically, chooses 
 whether to output the symbol $\sigma_{k}$ or to follow a backward
 link $S_{k}$. Another probabilistic version of this process can be found in
 \cite{pntcc}.

For this work, we have modeled $IMPROV$ as a simpler improvisation
 process. We are more interested in showing the synchronization
 between the improvisation phases, than showing how we can control the choice among  \textit{suffix links} and \textit{factor links} based on a probabilistic distribution. For that reason, choices in our $IMPROV$ process are made non-deterministically.\\

$IMPROV(k)$ $\overset{def}{=}$  \\
\hspace*{28pt}\textbf{when} $S_{k}=-1$ \textbf{do} \textbf{next} \\
\hspace*{35pt}(\textbf{tell} ($out = \sigma_{k+1}$) $\|$ $IMPROV(k+1)$)  \\
%\hspace*{31pt} \\
\hspace*{24pt}$\|$ \textbf{when} $S_{k} \geq 0$ \textbf{do}  \textbf{next} \\
\hspace*{40pt} ((\textbf{tell} ($out = \sigma_{k+1}$)  
%\\  \hspace*{42pt} 
$\|$ $IMPROV(k+1)$)  +\\
\hspace*{42pt}$\sum \limits_{\sigma \in \Sigma}$ \textbf{when}
$\sigma \in from_{s_{k}}$ \textbf{do} \\ 
\hspace*{82pt}( \textbf{tell} ($out = \sigma$)$\|$ $IMPROV(\delta_{s_{k}},\sigma)$))\\
\hspace*{23pt}$\|$ \textbf{unless} $S_{k} \geq -1$  \textbf{next} $IMPROV(k)$ \\

% We developed a $WAIT_{n}$ process to wait until $n$ symbols have been learned
% and launch the $IMPROV(k)$ process.

% $WAIT_{n}$ $\overset{def}{=}$   \textbf{when} $go = n$ \textbf{do}
% $IMPROV(n)$ \\ 
% \hspace*{50pt} $\|$ \textbf{unless} $go = n$ \textbf{do} \textbf{next} $WAIT_{n}$\\

The system is modeled as the $PLAYER$ and the $SYNC$ process running
in parallel with a process waiting until $n$
symbols have been played to launch the $IMPROV$ process.\\

$SYSTEM_{n}$ $\overset{def}{=}$  
 !\textbf{tell}($S_{0} = -1$) $\|$ $PLAYER_{1}$ \\
\hspace*{67pt}$\|$ $SYNC_{1}$ $\|$ $Wait_{n}$ 

\subsection{Signal processing}

\texttt{Ntcc} was used in the past as an audio processing framework
\cite{audiontcc}. In that work, Valencia and Rueda showed how this
modeling formalism gives a compact and precise definition of audio
stream systems. They argued that it is possible to model an audio
system and prove temporal properties using the temporal logic
associated to \texttt{ntcc}. They proposed that a \texttt{ntcc} model, where each time-unit can be associated to processing the current sample of a sequential stream.

Unfortunately, in practice it is difficult to implement that model because it will require to execute 44100 time-units per second to process a 44.1 kHz audio stream. This is not possible using our interpreter
and using the other \texttt{ntcc} interpreters neither.
%  Additionally, there is not an implementation of an efficient and generic constraint system for real numbers.

Another approach to give formal semantics to audio processing is the
visual audio processing language $Faust$ \cite{faust}. Faust semantics are
based on an algebra of block diagrams. This gives a formal and precise
meaning to the operation programed there. 
%$Faust$ has also
%been been interfaced with Pd \cite{AG07}.

Our approach is different, we use a \textit{Ntccrt} external for Pd or Max to synchronize the graphical objects
 in charge of audio, video or MIDI processing in Pd.
For instance, the \texttt{ntcc} external decides when triggering a graphical object in charge of applying a delay filter to an
audio stream and it will not allow other graphical objects to apply a filter on that audio stream, until the delay filter
finishes its work.

Our system is composed by a collection of
$n$ filters and $m$ objects
(MIDI, audio or video streams). 
%Some processes work on a single
%object, but others work on both objects. 
When a filter $P_i$ is
working on an object $m_j$, another filter cannot work on $m_j$ until
$P_i$ is done. A filter $P_i$ is activated when a condition over its input 
is true. That condition is easily represented by a constraint.

Our system is composed by the infinite rational tree variables $work$, $end$ and $input$ representing lists. $Work_j$ represents the identifiers of the filter working on the object $j$. $End_j$ represents when
the object $j$ has finished its work. Values for $end_j$ are updated each time-unit with information from the environment.
$Input_i$ represents the conditions necessary to launch filter $P_i$, based on information received from the environment.
Finally, $wait_j$ represents the set of filters waiting to work on the object $m_j$. Note that $work_j$ is a reference to the position $j$ of the list $work$ (same with $end$ and $input$).

Next, we explain the definitions of our system. Objects are represented by  \textit{IdleObject} and \textit{BusyObject}. An object is $idle$ until it
 non-deterministically chooses a filter from the $wait_j$ variable. After that, it will remain 
$busy$ until the $end_j = true$  constraint can be deduced from
the store. \\

$IdleObject(j)$ $\overset{def}{=}$ \\
\hspace*{27pt} \textbf{when} $work_j > 0$ \textbf{do} \textbf{next} $BusyObject(j)$  \\
\hspace*{21pt} $\|$ \textbf{unless} $work_j > 0$ \textbf{next} $IdleObject(j)$    \\
\hspace*{21pt} $\|$ $\sum \limits_{x \in P}$ \textbf{when} $x \in wait_j$ \textbf{do} \textbf{tell} $work_j = x$  \\

$BusyObject(j)$ $\overset{def}{=}$ \\
\hspace*{27pt} \textbf{when} $end_j = true$ \textbf{do} $IdleObject(j)$  \\
\hspace*{21pt} $\|$ \textbf{unless} $end_j = true$ \textbf{next} $BusyObject(j)$  \\

Filters are represented by the definitions \textit{IdleFilter}, \textit{WaitingFilter} and \textit{BusyFilter}. 
A filter is $idle$ until it can deduce that $input_i = true$. $Input_i$ can be a condition based on multiple criteria. \\

$IdleFilter(i,j)$ $\overset{def}{=}$ \\
\hspace*{24pt} \textbf{when} $input_i = true$ \textbf{do} $WaitFilter(i,j)$ \\
\hspace*{19pt} $\|$  \textbf{unless} $input_i = true$ \textbf{next} $IdleFilter(i,j)$ \\

A filter is $waiting$ when the information for launching it
can be deduced from the store, but it has not yet control over the object $m_j$. When it can control the object,
it calls the definition \textit{BusyFilter}. \\

$WaitingFilter(i,j)$ $\overset{def}{=}$ \\
\hspace*{24pt} \textbf{when} $work_j = i$ \textbf{do} $BusyFilter(i,j)$ \\
\hspace*{19pt} $\|$  \textbf{unless} $work_j = i$ \textbf{next} \\
\hspace*{35pt} $WaitingFilter(i,j)$ $\|$ \textbf{tell} $i \in wait_j$\\

A filter is $busy$ until it can deduce that the filter finished working
on the object associated to it. \\ %problem con la persistence de los ends

$BusyFilter(i,j)$ $\overset{def}{=}$ \\
\hspace*{24pt} \textbf{when} $end_j = true$ \textbf{do} $IdleFilter(i,j)$ \\
\hspace*{19pt} $\|$  \textbf{unless} $end_j = true$ \textbf{next} $BusyFilter(i,j)$ \\

\begin{figure}[htbp]
\centerline{\framebox{
	\includegraphics[width=0.9\columnwidth]{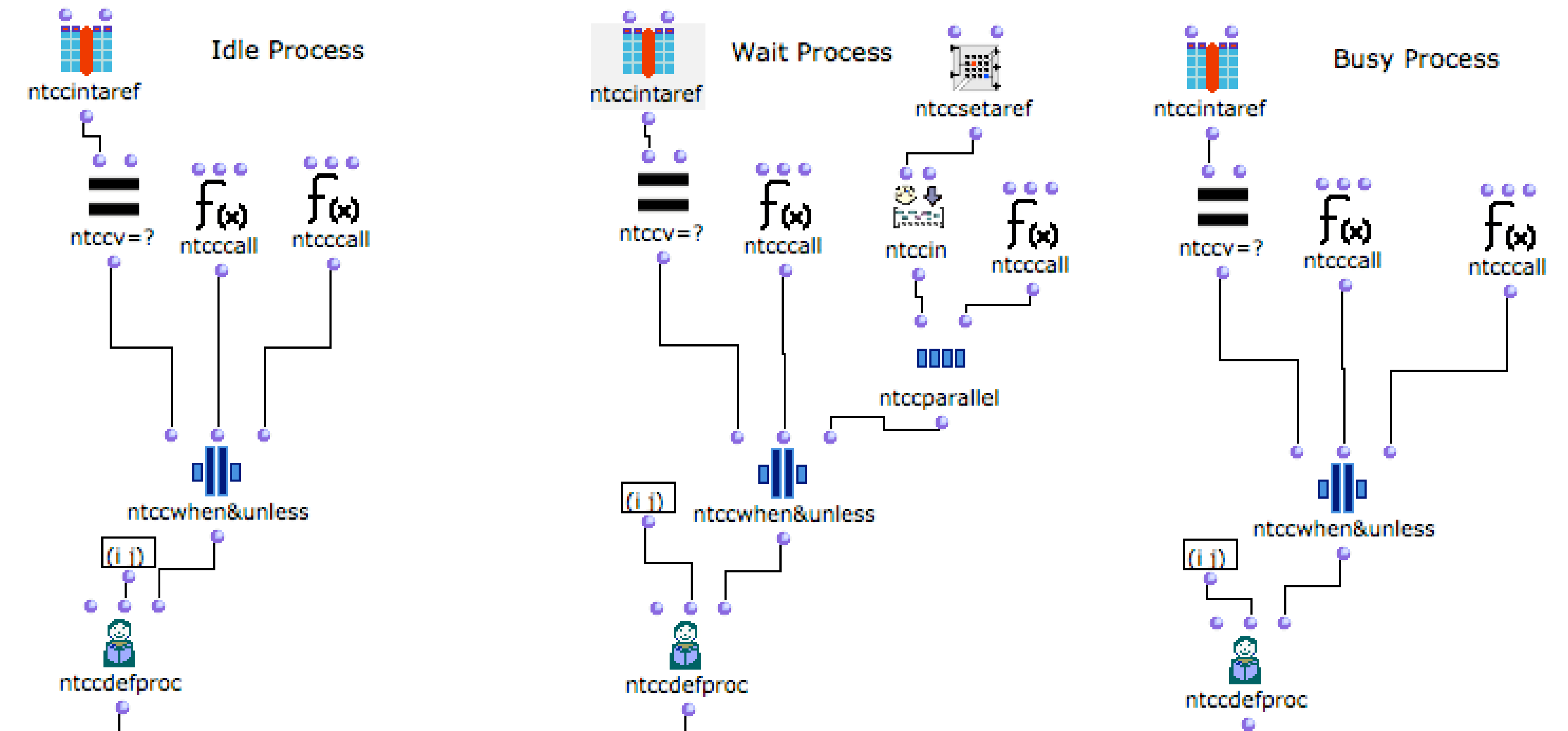}}  } 
\caption{Specifying a $Ntccrt$ external in OpenMusic.}
\label{fig:philo2}
\end{figure}

Filter definitions can be written in OpenMusic using a graphical ``patch'' (fig \ref{fig:philo2}).

%\hspace*{36pt} \textbf{when} $end_j = true \wedge (k = -1 \vee end_k = true)$ \textbf{do} $IdleProcess(i,j,k)$ \\
%\hspace*{31pt} $\|$ \textbf{unless} $end_j = true \wedge (k = -1 \vee end_k = true)$ \textbf{next} $BusyProcess(i,j,k)$  \\
%$end_j = true$ and $end_k = true$.  \\ 

The following definition models a situation with two objects and four filters. 
The external generated for this model can control all kind of objects and filters, represented
by graphical objects in Pd. \\

% The system only triggers the execution
%of each filter by posting the constraint $work_j = i$, receives a value for %when the filter is done, and receives another value for $input_i$ when the conditions to execute the filter $P_i$ are satisfied.\\

$System()$ $\overset{def}{=}$ \\
\hspace*{26pt} $IdleObject(1)$ $\|$ $IdleObject(2)$ $\|$ $IdleFilter(1,1)$ \\
\hspace*{18pt} $\|$ $IdleFilter(1,2)$ $\|$ $IdleFilter(2,1)$ $\|$ $IdleFilter(2,2)$\\
%hspace*{27pt}  \\  

%\section{\textit{Ntccrt} framework, screenshots and examples}

\section{Results}
% Using \textit{Ntccrt}, we ran $Ccfomi$. This was the first time, a 
% \texttt{ntcc} simulation of a multimedia interaction interacted with a human user in real-time. 

We ran $Ccfomi$ as an stand-alone application over an Intel 2.8 GHz iMac using
Mac OS 10.5.2 and Gecode 2.2.0. Each time-unit took an average of 20 ms, scheduling
around 880 processes per time-unit. We simulated 300 time-units and we ran each simulation 100 times in our tests.

% ``John  McLaughlin,  said  to  be  one  of  the  fastest  Jazz 
% guitarists,  and  found a minimum  inter onset  time of about 60 milliseconds. This  figure gives 
% an  approximate  constraint  for  the  computation  time of our  system:  it  should  be  able  to  learn 
% and produce  sequences  in  less  than  30  milliseconds.'' [The continuator]
Pachet argues in \cite{continuator} that an improvisation system able to learn and produce sequences in less than 30ms is appropriate for real-time interaction. Since our implementation of \textit{Ccfomi} has a response time of 20ms in average, we conclude that it is capable of real-time interaction for a 300 (or less) time-units simulation.

For this work, we made all the test under Mac OS X using Pd. Since we are using Gecode
and Flext to generate the externals, they could be easily compiled to other platforms
and for Max. This is due to Gecode and Flext portability.

\section{Conclusions and Future work}
In this paper, we present \textit{Ntccrt} as a framework to manage concurrency in Max and Pd. In addition, we present two real-life applications, a machine improvisation system and a signal processing system. We ran both applications creating \textit{Ntccrt} external objects for Pd. 

% If the reader does not consider relevant using process calculi (such as \texttt{ntcc}) to model, verify and execute a real-time system, we pose the  reader the following questions. Has the reader developed a real-time system, synchronizing processes based on complex conditions, on either Max or Pd? Try verifying the system formally! Is it an easy task? Would the reader be able to write an external to represent such system in 50 lines of code? Using \texttt{ntcc}, we did it.

We want to encourage the use of process calculi to develop reactive systems. For that reason, this research focuses on developing real-life applications with \texttt{ntcc} and showing that our interpreter \textit{Ntccrt} is a user-friend-ly tool, providing a graphical interface to specify \texttt{ntcc} models and compiling them to efficient C++ programs capable of real-time interaction in Pd. 

We argue that using process calculi (such as \texttt{ntcc}) to model, verify and execute reactive systems decreases the development time and guarantees correct process synchronization, in contrast to the graphical patch para-digm of Max or Pd. We argue that using that paradigm is difficult and time-demanding to synchronize processes depending on complex conditions. On the other hand, using Ntccrt, we can model such systems with a few graphical boxes in OpenMusic or with a few lines in Common Lisp, representing complex conditions by constraints.

% If one can model such systems using process calculi, 
% why they have not been used in real-life applications? Garavel argues in \cite{garavel} that models based on process calculi are not widespread because there are many calculi and many variants for each calculus, being difficult to choose the most appropriate. In addition, it is difficult to express an explicit notion of time and real-time requirements in process calculi. Finally, he argues that existing tools for process calculi are not user-friendly. 

% We want to make process calculi widespread for real-life applications. We left the task of representing real-time in process calculi and choosing the appropriate variant of each calculus for each application to senior researchers.

% We also showed that our approach to design \textit{Ntccrt} offers better performance than using threads or event-driven programming to represent the processes.

One may argue that although we can synchronize \textit{Ntccrt} with an external clock (e.g., a metronome object) provided by Max or Pd, this does not solve the problem of simulating models when the clock step is shorter than the time necessary to
compute a time-unit. To solve this problem, Sarria proposed to develop an interpreter for the Real Time Concurrent Constraint (\texttt{rtcc} \cite{rtcc}) calculus, which is an extension of \texttt{ntcc} capable of modeling \textit{time-units} with fixed duration. 

One may also argue that we encourage formal verification for \texttt{ntcc}, but there is not an existing tool to verify these models automatically, not even semi-automatically.
To solve this problem, P\'{e}rez and Rueda proposed to develop a verification tool for the Probabilistic Timed Concurrent Constraint (\texttt{pntcc} \cite{pntcc}) calculus. Currently, they are able to generate
an input for Prism \cite{prism08} based on a \texttt{pntcc} model. 

In the future, we would like to explore the ideas proposed by Sarria, P\'{e}rez and Rueda. Moreover, we want to extend our implementation to support \texttt{pntcc} and \texttt{rtcc}, and to generate an input for Spin \cite{spin} based on a \texttt{ntcc} model.

\section{Acknowledgments}
We want to thank to Arshia Cont for giving us this idea of using
\textit{Ntccrt} in Pd and Max; Fivos Maniatakos, Jorge P\'{e}rez and Carlos Toro-Berm\'{u}dez for their valuable reviews
on this paper; and Jean Bresson, Gustavo Guti\'{e}rrez, and the Gecode developers for their help during the development
of \textit{Ntccrt}.
%\balance

%\bibliographystyle{abbvr}
%\bibliography{mybib} % requires file template.bib

\let\oldbibliography\thebibliography
\renewcommand{\thebibliography}[1]{%
  \oldbibliography{#1}%
  \setlength{\itemsep}{0pt}%
}

\bibliographystyle{abbrv}
{\scriptsize
\bibliography{mybib}
}

\end{document}